\documentclass[superscriptaddress,reprint,aps,pra,longbibliography]{revtex4-1}
\usepackage{graphicx}
\usepackage{amsmath}
\usepackage{amsfonts}
\usepackage{enumerate}
\usepackage{multirow}
\usepackage{makecell}
\usepackage[table,xcdraw]{xcolor}
\usepackage{times}
\usepackage[normalem]{ulem}
\usepackage[hidelinks]{hyperref}
\hypersetup{
    colorlinks,
    linkcolor={blue!80!black},
    citecolor={blue!80!black},
    urlcolor={blue!80!black}}

\begin{document}

\title{Quantum synchronization in a collision model}

\author{G. Karpat}
\email{goktug.karpat@ieu.edu.tr}
\affiliation{Faculty of Arts and Sciences, Department of Physics, \.{I}zmir University of Economics, \.{I}zmir, 35330, Turkey}

\author{\.{I}. Yal\c{c}\i nkaya}
\affiliation{Department of Physics, Faculty of Nuclear Sciences and Physical Engineering, Czech Technical University in Prague, B\v{r}ehov\'a 7, 115 19 Praha 1-Star\'e M\v{e}sto, Czech Republic}

\author{B. \c{C}akmak}
\affiliation{College of Engineering and Natural Sciences, Bah\c{c}e\c{s}ehir University, Be\c{s}ikta\c{s}, \.{I}stanbul 34353, Turkey}

\date{\today}

\begin{abstract}
We reveal the emergence of environment-induced spontaneous synchronization between two spin-$1/2$ quantum objects in a collision model setting. In particular, we determine the conditions for the dynamical establishment of synchronous evolution between local spin observables of a pair of spins undergoing open-system dynamics in the absence of an external drive. Exploiting the versatility of the collision model framework, we show that the formation of quantum or classical correlations between the principal spin pair are of no significant relevance to the manifestation of spontaneous quantum synchronization between them. Furthermore, we discuss the consequences of thermal effects on the environmental spins for the emergence of quantum synchronization. 
\end{abstract}

\maketitle

\section{Introduction}

Synchronization is a ubiquitous phenomenon which is encountered in nature in various different contexts, ranging from sociology and biology to physics~\citep{Pikovsky2001,Osipov2007}. In fact, the first observation of synchronization dates back to 17th century when C. Huygens discovered an ``odd kind of sympathy'' between two pendulum clocks, realizing that their oscillations tend to exhibit synchronous behavior if they are coupled via a common support~\cite{Huygens1665}. Just to name a few examples, some other systems that show a tendency to operate in synchrony with each other include, but are not limited to, swinging metronomes, flashing fireflies, cardiac pacemakers and applauding audiences. In the case of forced synchronization, also known as entrainment, the system is driven by an external field acting as a pacemaker that tries to impose its rhythm on the system. On the other hand, the phenomenon of spontaneous synchronization manifests itself merely due to the coupling between two or more subsystems in the complete absence of an external drive. Despite the fact that the widespread concept of synchronization has been thoroughly investigated in the classical domain during the last few decades~\cite{Arenas2008}, its quantum mechanical counterpart has been considered only recently~\cite{Galve2017}.

In the quantum domain, the phenomenon of synchronization has been investigated under an external driving field in physical models such as spin-boson model~\cite{Goychuk2006,Ozgur2015}, qubits coupled to dissipative resonators~\cite{Zhirov2008,Zhirov2009}, and Van der Poll oscillators \citep{Lee2013,Walter2014,Sonar2018}. Regarding spontaneous quantum synchronization, numerous studies have also been performed in the last years, which for instance deal with system of harmonic oscillators~\cite{Giorgi2012,Manzano2013,Manzano2013a,Benedetti2016}, optomechanical arrays~\cite{Heinrich2011,Ludwig2013}, cold ions in microtraps \cite{Hush2015}, Van der Poll oscillators~\cite{Lee2014,Walter2015}, and spins interacting collectively with a common dissipative environment~\cite{Orth2010,Giorgi2013,Giorgi2016,Bellomo2017}. However, as compared to the harmonic and optomechanical systems, only a very limited number of studies has been carried out on the quantum synchronization of spin systems. While some of the recent works on this subject characterize quantum synchronization based on the requirements of the standard theory of synchronization in the quantum domain~\cite{Roulet2018,Roulet2018a,Jaseem2018}, others have followed a different approach and described quantum synchronization between the spins in terms of the synchronous behavior of local spin observables during a transient decay to the steady state~\cite{Giorgi2013,Giorgi2016}.

Spontaneous mutual synchronization emerges as result of some temporal correlation between the local evolution of the subsystems of the system of interest. Therefore, its signatures can be detected by analyzing the dynamical similarities between the local observables of the considered system. In this direction, different methods have been proposed to identify the appearance of quantum synchronization~\cite{Mari2013,Li2017}. Alternatively, one can look for the traces of synchronous behavior between local dynamics in the evolution of global correlations of the composite system. To this end, several measures of quantum and total correlations shared by the subsystems, e.g., entanglement, quantum discord, and mutual information, have been explored to pinpoint their relation to the emergence of mutual synchronization~\cite{Galve2017}. Even though behavior of certain global correlations between the subsystems and the occurrence of spontaneous synchronization have been observed to be related in some systems, a consistent connection between these two concepts has not been found in general.

Among numerous ways to simulate open quantum systems, collision models have recently attracted considerable attention due to their versatility in modeling different regimes of the dynamics, such as with or without memory~\cite{scarani2002,ziman2005a,ziman2005b,ziman2010,ciccarello2013,mccloskey2014,bernandes2014,vacchini2014,strunz2016,cakmak2017,ciccarello2017,lorenzo2017a,lorenzo2017b,filippov2017,jin2018,campbell2018,campbell2019b,cuevas2018}, which is much harder to implement with other approaches. In a collision model framework, the environment is constituted by a number of particles, which can be both finite dimensional and continuous variable, and the dynamics of the system is determined by its sequential interaction with these environmental units. Along with the described basic setting that yields a Markovian time evolution for the system, one can simply introduce intra-environment collisions to simulate a non-Markovian dynamics as well. Following the discussions on the memory effects and owing to their highly adoptable nature, quantum collision models have served as a test-bed for various ideas in several different fields, such as  quantum optics~\cite{ciccarello2017}, quantum thermodynamics~\cite{lorenzo2015a,lorenzo2015b,pezzutto2016,pezzutto2019,abah2019}, quantum control~\cite{cakmak2019,beyer2018}, and recently their all-optical experimental implementation has been demonstrated~\cite{cuevas2018}. 

In this work, we reveal the occurrence of spontaneous mutual synchronization between two spin-$1/2$ particles in a collision model framework. Making use of Pearson's coefficient to quantify the temporal correlations between the local dynamics of two spins forming the open quantum system, we determine the physical conditions under which expectation values of local spin observables become synchronized in time. Specifically, we identify the requirements for the establishment of synchronized dynamical behavior between the spin observables depending on the parameters of the model, namely, detuning between the two system spins, strength of the coupling between them, and the interaction strength among the environmental spins. Moreover, owing to the versatility of collision models which enables full control over the dynamics of two open system spins, we investigate the role of correlations between the spins for their synchronous behavior in detail. Our findings clearly demonstrate the indifference of the formation of correlations between the principal pair of spins for the emergence of spontaneous synchronization. Lastly, we study the consequences of having thermal effects on the environmental spins on the appearance of synchronization.

This manuscript is organized as follows. In Section~\ref{model}, we introduce the collision model and briefly discuss its characteristic properties. In Section~\ref{sync}, we specify the requirements for the emergence of spontaneous quantum synchronization in the considered model. We discuss the role of bipartite correlations between the open system spins, and thermal effects on the environmental spins, for the occurrence of synchronization. Section~\ref{conclusion} includes the summary of our results.

\section{Collision Model}\label{model}

\begin{figure}[t]
\centering
\includegraphics[width=.47\textwidth]{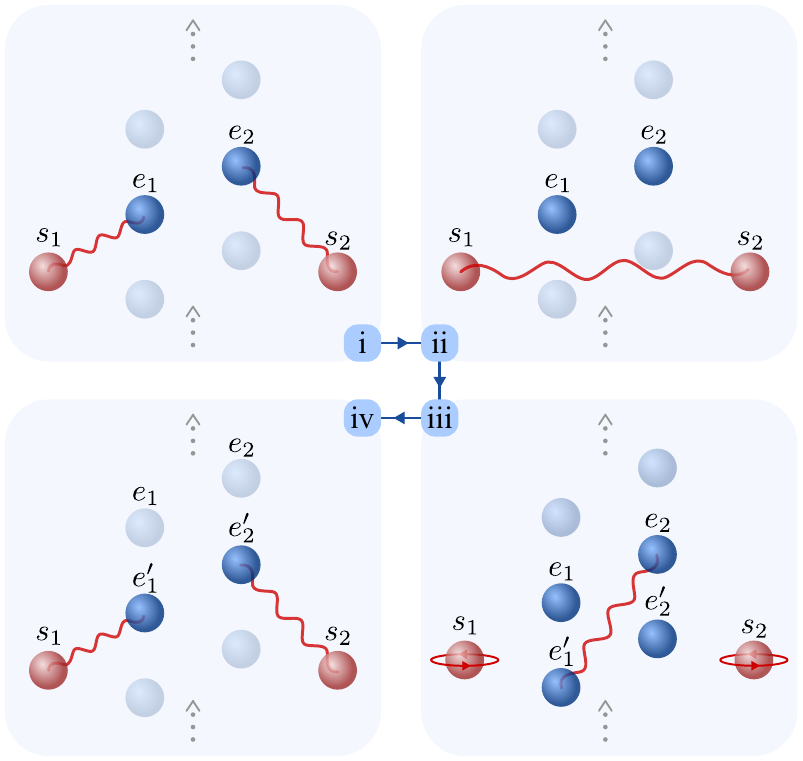}
\caption{Schematic sketch of the considered collision model. A single step in our model involves four adjacent collisions between the system $s$ and the environment $e$. (i) The system qubits, $s_1$ and $s_2$, interact with the environment qubits $e_1$ and $e_2$, respectively, that belong to their own environments. (ii) The system qubits directly interact with each other. (iii) The system qubits freely evolve while a partial SWAP interaction takes place between the environmental spins $e_2$ and $e_1’$. (iv) The environment spins which have interacted the system spins in step (i) are traced out. The system spins are now ready to couple with the forthcoming environmental spins $e_1'$ and $e_2'$.}
\label{fig1}
\end{figure}

In this section, we will describe the details of the collision model we intend to study in the rest of our analysis. Our system of interest is composed of two spin-$1/2$ particles, $s_1$ and $s_2$, that are in contact with their own environments, $e_1$ and $e_2$, separately. Particles constituting both environments are also spin-$1/2$ systems and, in principle, made out of infinite number of objects in the same initial state. It is assumed that there exist no initial correlations between the system or environment spins. Generally, in the collision model framework, the interactions between the system and environment take place as successive collisions, that is, as brief pairwise couplings described with the help of unitary evolution operators. As mentioned above, system spins are not allowed to interact with the environmental spins that belong to other system spin's environment, however, interactions between these environments are allowed to take place. After laying out these general rules for the model, we now present the specifics of a single step in the dynamics of our model in a point by point fashion, which is schematically summarized in Fig.~\ref{fig1}.
\begin{enumerate}[(i)]

\item The collision scheme begins with the interaction of both system particles $s_1$ and $s_2$ with their respective environments $e_1$ and $e_2$ via a XX-type coupling, which is described by the Hamiltonian
\begin{equation}
H_{se}= \frac{J}{2}(\sigma^{x}_{s} \sigma^{x}_{e} + \sigma^{y}_{s} \sigma^{y}_{e}),
\end{equation}
where $\sigma^{x}$ and $\sigma^{y}$ are the standard Pauli operators in the x- and y-directions, and the corresponding unitary evolution operators are written as $U_{se}=\exp(-i H_{se} \delta t_{se})$ (setting $\hbar=1$) with $J\delta t_{se}$ characterizing the strength of the coupling between the system and environment spins.

\item Then the system spins interact with each other through an Ising-like interaction Hamiltonian, i.e.,
\begin{equation}
H_{s_1s_2}= \frac{\lambda}{2} (\sigma^{x}_{s_1} \sigma^{x}_{s_2}),
\end{equation}
where the operator $U_{ss}=\exp(-i H_{s_1s_2} \delta _{t_{ss}})$ describes their evolution with $\lambda\delta t_{ss}$ being the interaction strength between the two system spins $s_1$ and $s_2$. 

\item In the next step, we let the system spins $s_1$ and $s_2$ evolve freely with their self-Hamiltonians 
\begin{eqnarray}
H_{s_{1}}=- \frac{\omega_1}{2} \sigma^{z}_{s_1}, \quad H_{s_{2}}= -\frac{\omega_2}{2} \sigma^{z}_{s_2},
\end{eqnarray}
where $\sigma^{z}$ is the Pauli operator in the z-direction. Here $\omega_1$ and $\omega_2$ represent the self energies of the system spins $s_1$ and $s_2$, respectively, with their self-evolution operators $U_{s_{1}}=\exp(-i H_{s_1} \delta t_s)$ and $U_{s_{2}}=\exp(-i H_{s_2} \delta t_s)$. Meanwhile, the environment spin $e_2$, which belongs to the environment of the second system spin $s_2$ and which has interacted with it in (i), gets coupled with the forthcoming environment spin $e^{\prime}_1$ that belongs to the environment of the first system spin $s_1$, through a partial SWAP operation given by
\begin{equation}\label{swap}
U_{e_{2}e^{\prime}_1}= \cos(\gamma)\mathbb{I}_4+i \sin(\gamma) \text{SWAP},
\end{equation}
where $\mathbb{I}_4$ denotes the $4\times4$ identity operator and the parameter $\gamma$ is the strength of the unitary partial SWAP operation with $\text{SWAP}=|00\rangle \langle00|+|01\rangle \langle10|+|10\rangle \langle01|+|11\rangle \langle11|$ written in the computational basis. We should note that $U_{e_2e_1'}$ in Eq. (\ref{swap}) can also be simply generated by a Heisenberg-type Hamiltonian.

\item Finally, we complete a single step of the presented collision scheme by discarding the environment spins $e_1$ and $e_2$, and move on to repeating the described procedure with the new environments $e^{\prime}_1$ and $e^{\prime}_2$. At this stage, a subtle point arises on whether or not to carry the correlations established between the system spins and the forthcoming environments to the next step of the dynamics. In what follows, we have considered both cases by employing two different trace out strategies:
\begin{enumerate}[(a)]
\item In passing from the $i^{th}$ step to $(i+1)^{th}$ step in the dynamics, we perform the  trace out strategy:
\begin{equation*}\label{rhocorr}
\rho_{s_1s_2e^{\prime}_1e^{\prime}_2}=\text{Tr}_{e_1e_2}[U \rho_{s_1s_2e_1e_2e^{\prime}_1e^{\prime}_2} U^\dagger], 
\end{equation*} 
where $U= U_{e_{2}e^{\prime}_1} U_{s_{2}}U_{s_{1}} U_{s_{1}s_{2}} U_{s_{2}e_{2}} U_{s_{1}e_{1}}$ and continue the $i+1^{th}$ step with the state above together with two fresh environmental spins. In this way, it is possible keep the correlations that have been established among the system spins $s_1$ and $s_2$, and the environment spins $e^{\prime}_1$ and $e^{\prime}_2$ throughout the dynamics as described in (i)-(iii).

\item Alternatively, we can adopt the following strategy:
\begin{align*}
\rho_{s_1} &=\text{Tr}_{s_2e_1e_2e^{\prime}_1e^{\prime}_2}[U \rho_{s_1s_2e_1e_2e^{\prime}_1e^{\prime}_2} U^\dagger], \\ 
\rho_{s_2} &=\text{Tr}_{s_1e_1e_2e^{\prime}_1e^{\prime}_2}[U \rho_{s_1s_2e_1e_2e^{\prime}_1e^{\prime}_2} U^\dagger], \\ 
\rho_{e^{\prime}_1e^{\prime}_2} &=\text{Tr}_{s_1s_2e_1e_2}[U \rho_{s_1s_2e_1e_2e^{\prime}_1e^{\prime}_2} U^\dagger], 
\end{align*}
and explicitly erase all sorts of correlations that have been formed between the system and environment spins before the next leg of collisions begin. As a result, we can start the forthcoming round with the initial state $\rho_{s_1}\otimes\rho_{s_2} \otimes \rho_{e^{\prime}_1e^{\prime}_2}$, once again together with two upcoming fresh environments.
\end{enumerate}
\end{enumerate}

{At this point, we would also like to mention that assuming the existence of an XX-type interaction between the system and environment spins in the first leg of the model is indeed a physically reasonable consideration, since it actually corresponds to the exchange interaction term in dipole-dipole type couplings, which is relevant for spins interacting in close proximity to each other. On the other hand, one might wonder the reason why we have considered a different interaction between the system spins $s_1$ and $s_2$ in the second leg of the collision scheme. In fact, we could have chosen the same XX-type interaction between the system spins as well, instead of the Ising interaction, but this produces qualitatively identical results for emergence of spontaneous synchronization. Hence, here we have considered the simplest possible interaction between the system spins for our purposes.} As a final remark, it is important to emphasize that while $s_2$ always interacts with a fresh environmental unit at every step of the dynamics, the environment spin with which $s_1$ interacts, gets partially swapped with the environment spin that has interacted with $s_2$, due to the partial SWAP interaction. This in turn implies an indirect interaction between $s_1$ and $s_2$ which is mediated through the coupling between the two separate environments.

\section{Quantum Synchronization}\label{sync}

Having introduced the essential features of the considered collision model, we are now in a position to commence our investigation on the spontaneous synchronization between the system spins $s_1$ and $s_2$. In the following, we present our results in three separate parts. First, we establish the physical conditions such that the expectation values for the local observables of the system spins $s_1$ and $s_2$ become spontaneously synchronized in time without any external drive on the system. Second, utilizing the aforementioned distinct trace out strategies, we show that the presence or absence of quantum or total correlations, in the form of entanglement and mutual information respectively, between the system spins $s_1$ and $s_2$ have no significant effect for the emergence of quantum synchronization. Finally, we demonstrate that thermal effects on the environment spins tend to degrade the synchronous behavior between the system spins. 

\subsection{Mutual synchronization}

We start our discussion noting that the analysis of spontaneous mutual synchronization presented in our work is independent of the initial state of the system spins $s_1$ and $s_2$ and the choice of local spin observable. That is, even though it is possible to obtain quantitatively different results by focusing on different system initial states and local spin observables, the qualitative outcomes of the analysis would remain unchanged. On the other hand, here we will assume that all the identical environment spins are initially in their ground state, i.e., $\rho_e=|0\rangle \langle0|$. Since the initial state of the environment spins affect the characteristics of the evolution due to the nature of the collision model, consequences of assuming environment spins in different initial states will be discussed particularly later on in a separate subsection. 

\begin{figure*}[t]
\centering
\includegraphics[width=.95\textwidth]{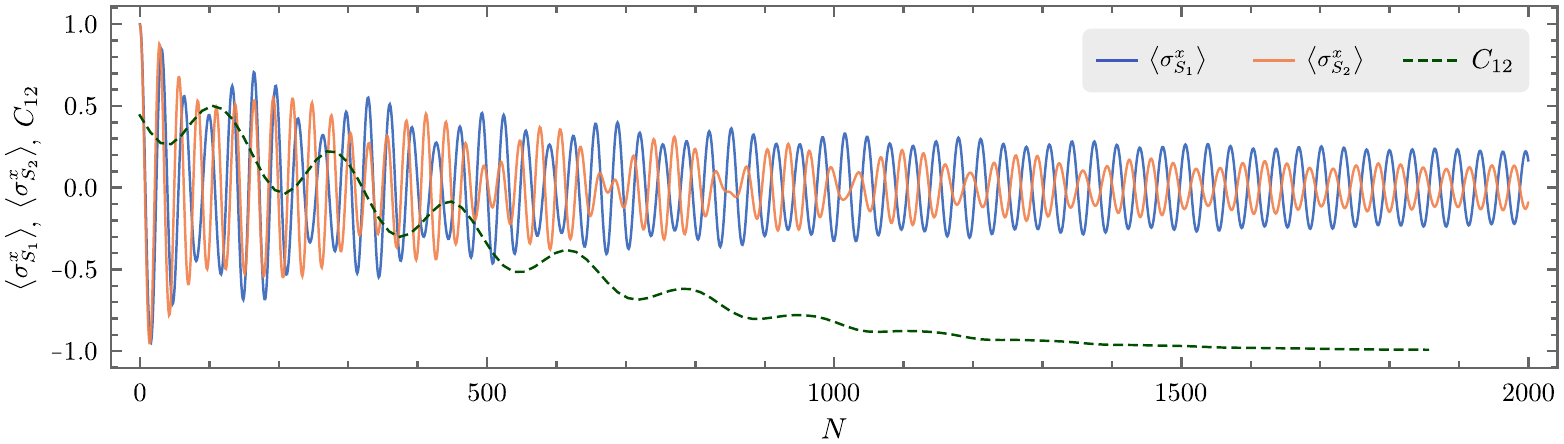}
\caption{Local spin expectation values for the system spins $s_1$ and $s_2$ along the x-direction and the Pearson correlation coefficient $C_{12}$, which is plotted for data windows of $140$ collisions with partial overlaps of $125$, versus the total number of collisions $N$. Anti-synchronization can be seen to be dynamically established after a certain number of collisions take place between the system and environment spins.}
\label{fig2}
\end{figure*}

In fact, the expectation value of any observable, expressed as a combination of the operators $\{\sigma^x,\sigma^y,\sigma^z,\mathbb{I}_2\}$, can be used to monitor the build-up of synchronization between the system spins $s_1$ and $s_2$ throughout the evolution. To be specific and precise, we intend to focus on the expectation value of the Pauli operator along x-direction, i.e., $\langle \sigma^x_{s} \rangle =\text{Tr}[\rho_s \sigma^x]$, to witness the dynamical emergence of synchronization. Besides, as we suppose that the pair of system spins are uncorrelated initially, their initial state can be in general written as $(\cos\theta_1 |0\rangle + e^{i \phi_1} \sin\theta_1 |1\rangle)\otimes(\cos\theta_2 |0\rangle + e^{i \phi_2}\sin \theta_2 |1\rangle)$. However, as the synchronizing nature of the dynamics is not affected by the choice of initial system states, we fix it as $(1/2)(|0\rangle+|1\rangle)\otimes(|0\rangle+|1\rangle)$ for concreteness.

The Pearson product-moment correlation coefficient $C_{12}$ can be used as a figure of merit to characterize the degree of synchronization between the expectation values of the local observables of system spins \cite{Galve2017}. Indeed, the Pearson coefficient is employed to measure the linear association between two discrete variables $x$ and $y$, and it is given by
\begin{equation}
C_{12}=\frac{\sum_{i=1}^n (x_i - \bar{x})(y_i - \bar{y})}{\sqrt{\sum_{i=1}^n (x_i - \bar{x})^2 \sum_{i=1}^n (y_i - \bar{y})^2}},
\label{eq:pearsonCoeff}
\end{equation}
where $\bar{x}$ and $\bar{y}$ are the mean values of $x$ and $y$, respectively, and $n$ is the number of values that the variables can assume. The Pearson coefficient $C_{12}$ can take a range of values between -1 and 1. While a vanishing value of $C_{12}$ indicates that the two variables are completely uncorrelated, a full positive association ($C_{12}=1$) means that as the value of one variable changes, so does the value of the other variable in the same way. On the other hand, a full negative association ($C_{12}=-1$) points out that the variables show anti-correlated behavior, that is, whereas the value of one variable increases, the value of the other variable decreases.  In our work, the variables under study are the local expectation values of the system spins along the x-direction, namely, $\langle \sigma^x_{s_1} \rangle$ and $\langle \sigma^x_{s_2} \rangle$, which both depend on the number of collisions $N$. Therefore, while the value of $C_{12}$ tend to zero for uncorrelated oscillations of the expectation values, the phase-locked oscillations between them are signaled when $|C_{12}|\approx 1$. In particular, mutual synchronization is said to be established if $C_{12}\approx 1$ and anti-synchronization emerges when $C_{12}\approx -1$. We are mostly interested in how our system switches {between} these cases  as it goes through more and more collisions. To this end, the summations appear in Eq. \eqref{eq:pearsonCoeff} are taken over not the whole but a part of the data set, called the data window whose width is $n$. In this way, the relation between $C_{12}$ and $N$ is obtained by evaluating all $C_{12}$'s as the data window slides along the whole data set. The overlap between adjacent data windows determines the smoothness and the resolution of the result. In our calculations, we choose $n$ to cover several oscillations of the expectation values together with a considerable overlap to avoid sharp fluctuations in $C_{12}$. {We provide a detailed elaboration of the evaluation of the Pearson coefficient in our work in Appendix \ref{appendix:A}}.

In this subsection, we will be hereafter examining the collision model with the trace out strategy (a), which has been explained in step (iv) of Sec.~\ref{model}. In other words, we will keep the correlations established among the system spins $s_1$ and $s_2$, and the environment spins $e^{\prime}_1$ and $e^{\prime}_2$ after each step of the described collision model. We consider an identical interaction between the system spins $s_1$ and $s_2$ and their respective environmental spins $e_1$ and $e_2$ setting $J\delta t_{se}=0.05$. The strength of the coupling between the system spins $s_1$ and $s_2$ is set as $\lambda\delta t_{ss}=0.03$, where the self energies of them are respectively fixed as $\omega_1=1$ and $\omega_2=1.1$ with $\delta t_{s}=0.2$. Here we assume that a strong partial SWAP operation takes place between the environmental spins $e_2$ and $e^{\prime}_1$, that is, we set the strength of the SWAP operation as $\gamma=0.95 \pi/2$. Taking into account the above specified parameters, Fig.~\ref{fig2} displays the evolution of the expectation values of local spin observables $\langle \sigma^x_{s_1} \rangle${with blue (dark gray) line} and $\langle \sigma^x_{s_2} \rangle$  {with orange (light gray) line} for 2000 collisions. In the same figure, along with the expectation values, we also show the Pearson coefficient $C_{12}$ {with dark green (dashed) line} which has been calculated for data windows of $140$ collisions with partial overlaps of $125$. In fact, it is not difficult to see that after a certain collision interval of incoherent oscillations, the spin expectation values become phase-locked as witnessed by the  behavior of the Pearson correlation coefficient. Therefore, Fig.~\ref{fig2} clearly demonstrates an example for the dynamical establishment of environment-induced spontaneous mutual anti-synchronization in the complete absence of an external drive in a quantum collision model.

\begin{figure*}[t]
\centering
\includegraphics[width=0.77\textwidth]{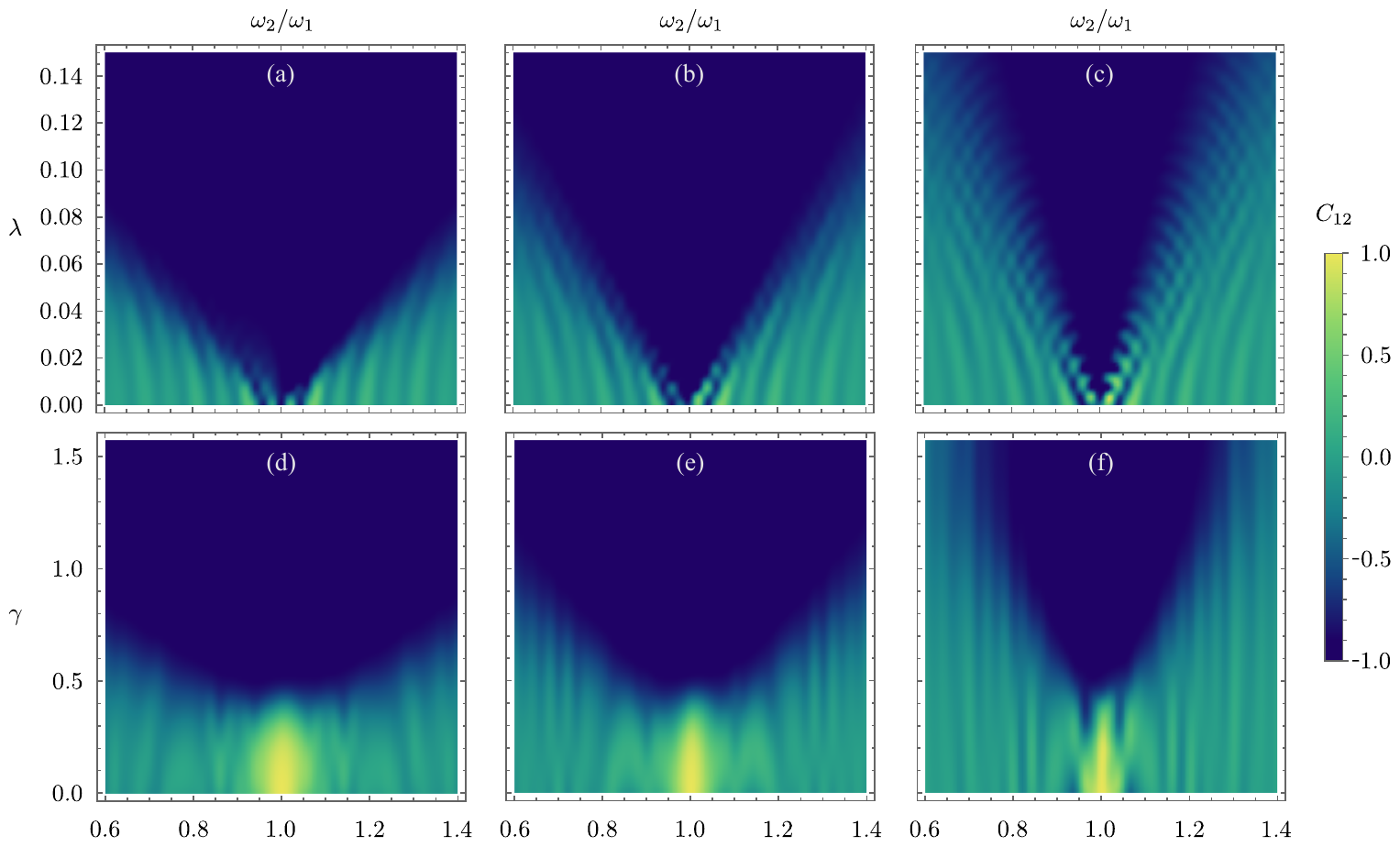}
\caption{Synchronization diagrams for different parameter regions. The Pearson correlation coefficient $C_{12}$ is evaluated for the local spin expectation values $\langle \sigma^x_{s_1} \rangle$ and $\langle \sigma^x_{s_2} \rangle$ for the system spins $s_1$ and $s_2$ for data windows of $250$ collisions with overlaps of 200. In the diagrams (a)-(c), the final value of $C_{12}$ after 6000 successive collisions is plotted versus the interaction strength between the system spins $\lambda$ and the ratio of the self-energies of the system spins $\omega_2/\omega_1$, for decreasing values of the interaction strength $\gamma$ between the environments $e_2$ and $e^{\prime}_1$ from left to right. The diagrams in (d)-(f) on the other hand show the final value of $C_{12}$ after 6000 successive collisions in terms of the strength of the partial SWAP operation $\gamma$ between the environment spins $e_2$ and $e^{\prime}_1$ and the ratio of the self-energies of the system spins $\omega_2/\omega_1$, for decreasing values of the coupling strength $\lambda$ between the system spins $s_1$ and  $s_2$ from left to right.}
\label{fig3}
\end{figure*}

Having witnessed above an instance of spontaneous synchronization in the collision model setting for a certain set of model parameters, we now turn our attention to a more thorough understanding of this phenomenon. An in-depth characterization of the dynamical manifestation of spontaneous mutual anti-synchronization between the pair of system spins $s_1$ and $s_2$ is presented in Fig.~\ref{fig3}, where we have once again evaluated the Pearson coefficient $C_{12}$ for the local spin observables $\langle \sigma^x_{s_1} \rangle$ and $\langle \sigma^x_{s_2} \rangle$. Here we have simulated a total number of 6000 collisions and $C_{12}$ has been calculated for data windows of $250$ collisions with partial overlaps of $200$. Indeed, in the presented plots, we display the final value of the Pearson coefficient $C_{12}$ with a color-coded legend after 6000 collisions take place between the system and the environment spins. Also, as in the example of Fig.~\ref{fig2}, we assume an identical interaction between the system spins $s_1$ and $s_2$ and their respective environment spins $e_1$ and $e_2$ setting $J\delta t_{se}=0.05$.

In Fig.~\ref{fig3} (a)-(c), we show three synchronization diagrams in terms of the coupling strength $\lambda$ ($\delta t_{ss}=0.2$) between the system spins $s_1$ and $s_2$, and the ratio of the self-energies of the system spins $\omega_2/\omega_1$ ($\omega_1=1$ and $\delta t_s=0.2$) for three distinct values of the partial SWAP interaction strength between the environment spins $e_2$ and $e^{\prime}_1$, namely, for $\gamma=0.95 \pi/2$ in (a), $\gamma=0.60 \pi/2$ in (b), and $\gamma=0.40 \pi/2$ in (c).
{To begin with, we would like to stress that, for all strengths of interaction between the environment spins considered in these three plots, spontaneous anti-synchronization always emerges, even in the complete absence of the coupling between the system spins, that is, $\lambda=0$, when there is no detuning between the self-energies of the system spins $s_1$ and $s_2$ , i.e. $\omega_1=\omega_2$. {(See Appendix \ref{appendix:B})}. Moreover, in each of one these plots, we can observe that the larger the amount of detuning, the stronger the interaction between the system spins should be for the onset of spontaneous anti-synchronization.} {Therefore}, from the point of view of the emergence of anti-synchronization, the main role of the interaction between the system spins $s_1$ and $s_2$ is merely to compensate for the detuning between the self energies, {which in indeed in line with some other studies in the literature, such as Ref. \citep{Giorgi2012}.} To put it differently, if the coupling between the pair of system spins is not sufficiently strong as compared to the degree of detuning between them, then anti-synchronization cannot be fully established between the local spin observables. In addition, the fact that the interaction between the environment spins $e_2$ and $e^{\prime}_1$ through the partial SWAP operation gets weaker from (a) to (c) results in a progressively narrower region of anti-synchronization.

In order to better comprehend the reason behind such a behavior, we present three more synchronization diagrams in Fig.~\ref{fig3} (d)-(f) in terms of the strength of the partial SWAP operation $\gamma$ between the environment spins $e_2$ and $e^{\prime}_1$, and the ratio of the self-energies of the system spins $\omega_2/\omega_1$ ($\omega_1=1$ and $\delta t_s=0.2$) for three values of the coupling strength between them, i.e., for $\lambda\delta t_{ss}=0.03$ in (d), $\lambda\delta t_{ss}=0.02$ in (e), and $\lambda\delta t_{ss}=0.01$ in (f). It is evident from these plots that the strength of the interaction $\gamma$ between the environment spins $e_2$ and $e^{\prime}_1$, which is in fact responsible for the indirect coupling between the system spins $s_1$ and $s_2$, must exceed some threshold values for the spontaneous anti-synchronization to occur between the local spin observables $\langle \sigma^x_{s_1} \rangle$ and $\langle \sigma^x_{s_2} \rangle$ after 6000 successive collisions. We should also emphasize that the small yellow regions near $w_1\approx w_2$ in these three plots, where the Pearson correlation coefficient $C_{12}\approx1$, are a consequence of the insufficient degree of information transfer between the environments for the establishment of anti-synchronization between the pair of system spins. Indeed, these yellow regions in Fig.~\ref{fig3} (d)-(f) are actually nothing other than the traces of the trivial synchronous dynamics between the local observables of the system spins due to our choice of identical initial system states. {It is crucial to emphasize here that without the presence of a sufficiently significant interaction between the environment spins, anti-synchronization cannot be established between the observables of the system spins, independent of the strength of coupling between the system spins.} {One may refer to Appendix \ref{appendix:B} for the detailed justification of this point.}

To summarize our findings in this part, it can be said that the establishment of environment-induced spontaneous synchronization crucially depends on the convey of information among the environment spins. In fact, this gives rise to an indirect transfer of information between the system spins, which is what  makes spontaneous mutual synchronization between them possible. Additionally, it can also be seen that the main task of the direct interaction between the system spins is solely the compensation of the detuning between their self energies.

\subsection{Role of correlations}

\begin{figure}[t]
\centering
\includegraphics[scale=1.75]{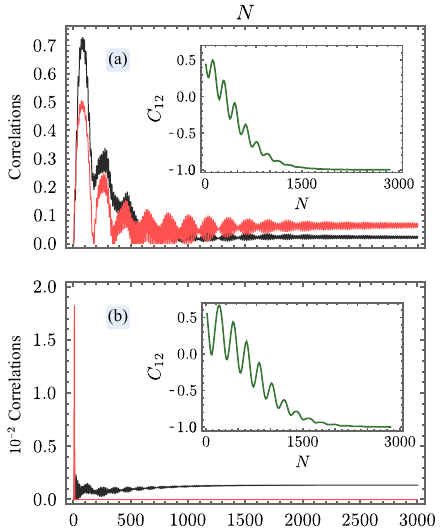}
\caption{Quantum and total correlations as measured by the concurrence $E$ {shown by red (light gray) line} and the quantum mutual information $I$ {shown by black line} versus the total number of collisions $N$. While (a) demonstrates the behavior of entanglement and mutual information assuming we keep the correlations among the system and environment spins after each leg of collisions, (b) shows their behavior in case we erase the established correlations after each leg of collisions. In the insets, we display the evolution of the Pearson correlation coefficient $C_{12}$ for data windows of $140$ collisions with partial overlaps of $125$.}
\label{fig4}
\end{figure}

In this subsection, we explore the potential consequences of the formation of quantum or total correlations between the system spins, throughout the collision dynamics, on the spontaneous emergence of synchronization between the local observables of system spins. Despite the fact that considerable effort has been expended in recent literature to comprehend the relation of the manifestation of synchronous dynamics between two quantum objects and the existence of correlations shared by them (see, for instance, Refs.~\citep{Galve2017,Ameri2015,Manzano2013,Giorgi2012,Giorgi2013}), no such general connection could be shown to exist, which is especially true in case of spin systems. Therefore, motivated by the versatility of the collision models, that is, by the ability to fully control the discrete evolution of the system and environment spins, we intend to understand whether such a correspondence exists between the phenomenon of synchronization and the formation of quantum and total correlations, quantified via entanglement and mutual information, respectively. In our analysis, we measure the degree of entanglement in the bipartite state of system spins $\rho_{s_{1}s_{2}}$ making use of concurrence~\cite{Wooters1998}. To evaluate concurrence, we need to  calculate the spin-flipped density matrix $\tilde{\rho}_{s_{1}s_{2}}=(\sigma^{y}\otimes\sigma^{y})\rho_{s_{1}s_{2}}^{*}(\sigma^{y}\otimes\sigma^{y})$, where $\rho_{s_{1}s_{2}}^{*}$ is obtained from $\rho_{s_{1}s_{2}}$ via complex conjugation. Then, the amount of entanglement in the state $\rho_{s_{1}s_{2}}$ reads
\begin{equation}
E(\rho_{s_{1}s_{2}})=\max \left\{ 0,\sqrt{\lambda_{1}}-\sqrt{\lambda_{2}}-\sqrt{\lambda_{3}}-\sqrt{\lambda_{4}}\right\},
\end{equation}
where $\{\lambda_{i}\}$ are the eigenvalues of $\rho_{s_{1}s_{2}} \tilde{\rho}_{s_{1}s_{2}}$ in decreasing order. On the other hand, we quantify the total amount of classical and quantum correlations in the bipartite system state $\rho_{s_{1}s_{2}}$ using quantum mutual information~\cite{Cerf1998}
\begin{align}
I(\rho_{s_{1}s_{2}})=S(\rho_{s_{1}})+S(\rho_{s_{2}})-S(\rho_{s_{1}s_{2}}),
\end{align}
where $S(\rho)=-\text{Tr}[\rho\log\rho]$ is the von Neumann entropy.  

In order to analyze the relationship between the bipartite system correlations and the emergence of synchronization, we will refer to the two distinct strategies that have been discussed in step (iv) of Sec.~\ref{model}. We first recall that following the first trace-out strategy (a) lets us keep the correlations, which have been established among the system and environment spins, intact during the time evolution of the model. Contrarily, in case of the second trace-out strategy (b) all the dynamically established correlations are explicitly erased after each step of collision dynamics is completed. In Fig.~\ref{fig4}, we present the dynamics of the concurrence $E$ {with red (light gray) line} and the quantum mutual information $I$ {with black line} as a function of the number of collisions, both in case of keeping (a) and erasing (b) the bipartite correlations between the system spins $s_{1}$ and $s_{2}$ after each leg of successive collisions, i.e., following the first and the second trace-out strategies described earlier. In addition, along with the correlations, we also show the evolution of the Pearson coefficient $C_{12}$ in the insets of both Fig.~\ref{fig4} (a) and (b) for data windows of $140$ collisions with partial overlaps of $125$. To obtain both plots here, we have supposed that system spins interact with their environment spins identically with $J\delta t_{se}=0.05$ and the coupling strength between the system spins are given by $\lambda\delta t_{ss}=0.03$, where the self energies of them are $\omega_1=1$ and $\omega_2=1.1$ with $\delta t_{s}=0.2$. Also, the interactions between the subsequent environment spins are characterized by $\gamma=0.95\pi/2$. Observing the plots (a) and (b) with their insets in Fig.~\ref{fig4}, it is straightforward to notice that the existence of quantum or total correlations throughout the collision dynamics is not of grave importance for the manifestation of synchronization between the local spin observables of the system spins $s_1$ and $s_2$. As a matter of fact, as demonstrated by the behavior of the Pearson correlation coefficient $C_{12}$ in the insets of the figure, anti-synchronization is established slightly quicker when we keep the correlations intact after each leg of successive collisions, as compared to the case of explicitly erasing the correlations after each leg of the collision scheme. All the same, it is still not difficult to perceive that the development or absence of quantum synchronization here is independent of the existence of quantum or total correlations between the system spins. 

\begin{figure}[t]
\centering
\includegraphics[scale=1.13]{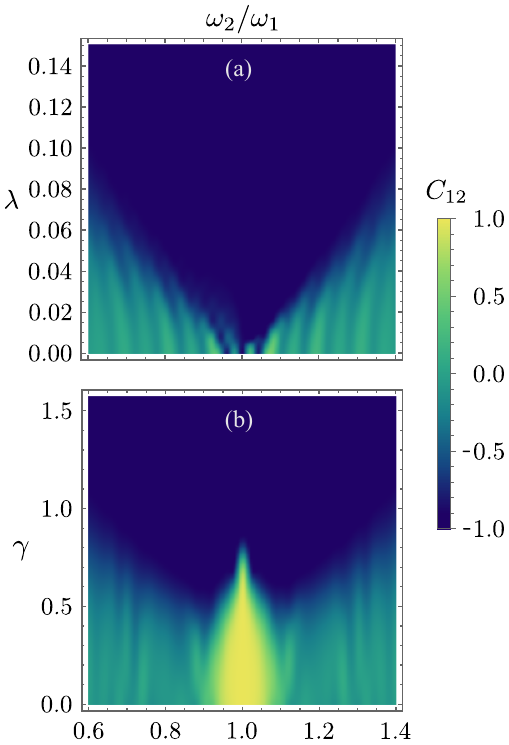}
\caption{Synchronization diagrams that are obtained via the second trace-out strategy (b) in which correlations among the system and the environment spins are erased after each leg of collisions. The Pearson coefficient $C_{12}$ is evaluated for $\langle \sigma^x_{s_1} \rangle$ and $\langle \sigma^x_{s_2} \rangle$ for data windows of $250$ collisions with partial overlaps $200$. In (a) the final value of $C_{12}$ after 6000 collisions is plotted versus the coupling strength $\lambda$ between the system spins supposing that $\gamma=0.95\pi/2$. In (b) the final value of $C_{12}$ is plotted versus the interaction strength between the environment spins assuming that $\lambda\delta t_{ss}=0.03$.}
\label{fig5}
\end{figure}

Since the above discussion holds only for a specific set of model parameters, it might not be sufficient to reflect the general relation between the formation of correlations and the establishment of spontaneous synchronization. Therefore, in Fig.~\ref{fig5} we present two other synchronization diagrams, similar to the ones shown in Fig.~\ref{fig3}, but this time adopting the second trace-out method in step (iv) of Sec.~\ref{model} where the correlations are not preserved after successive collisions. In particular, the plots in Fig.~\ref{fig5} show the final value of the Pearson correlation coefficient $C_{12}$ for the local spin observables $\langle \sigma^x_{s_1} \rangle$ and $\langle \sigma^x_{s_2} \rangle$ after 6000 successive collisions take place for data windows of $250$ collisions with partial overlaps of $200$. We also note that to obtain the results shown in Fig.~\ref{fig5} (a) and (b), we have used the exact same set of model and initial state parameters that have been utilized for obtaining Fig.~\ref{fig3} (a) and (d) respectively. Specifically, Fig.~\ref{fig5} (a) exposes how the establishment of anti-synchronization depends on the coupling strength between the system spins $s_1$ and $s_2$, and the ratio of the self energies of the system spins when we have a strong partial SWAP interaction between the environment spins $e_2$ and $e^{\prime}_1$ given by $\gamma=0.95 \pi/2$. On the other hand, Fig.~\ref{fig5} (b) shows the degree of spontaneous anti-synchronization as a function of the strength of the SWAP interaction between the environment spins and the ratio of the self energies of the system spins in case we have $\lambda \delta t_{ss}=0.03$. Consequently, comparing the results displayed in Fig.~\ref{fig3} (a) and (d) with the outcomes of Fig.~\ref{fig5} (a) and (b), it can be readily deduced that the existence of bipartite quantum or total correlations between the system spins $s_1$ and $s_2$ throughout the collision dynamics is of no significant relevance for the emergence of the spontaneous anti-synchronization provided that the interaction between the environment spins $e_2$ and $e^{\prime}_1$ is sufficiently strong. {We also} note that even though we have considered concurrence and mutual information to quantify the system correlations, the conclusions drawn from this section should in fact hold for any correlation measure since we erase all types of correlations between the system spins in the second trace-out strategy. {In concluding this subsection, we would like to lastly mention that as we discard the environment particles after each collision cycle by tracing them out due to the nature of collision models, it is also not possible for the correlations between the environment spins to play any role in the emergence of spontaneous anti-synchronization in our analysis.}

\subsection{Thermal effects}

Having characterized the spontaneous occurrence of synchronization between the system spins $s_1$ and $s_2$, in relation with the formation of bipartite quantum and total correlations shared by them, we will next consider a generalization of the so far discussed collision model. Particularly, rather than assuming that the environment spins $e_1$ and $e_2$ are initially in their identical ground state $\rho_e=|0\rangle\langle0|$, we now suppose that they are thermalized with two separate reservoirs at temperatures $T_1$ and $T_2$, respectively. That is to say that we take the initial state of the environment spins to be 
\begin{equation}
\rho_{e_{1}}=\frac{e^{-\beta_1 \omega_1 \sigma^z}}{\text{Tr}[e^{-\beta_1 \omega_1 \sigma^z}]}, \quad
\rho_{e_{2}}=\frac{e^{-\beta_2 \omega_2 \sigma^z}}{\text{Tr}[e^{-\beta_2 \omega_2 \sigma^z}]},
\end{equation}
where $\beta_{1(2)}=1/T_{1(2)}$ and the Boltzmann constant $k$ is set to unity. We should note that in the zero temperature limit $T=0$, we recover the previously examined the case, where all the environment spins are identical and initially in their ground state. On the other hand, when the equilibrium temperature tends to infinity as $T \rightarrow \infty$, all the environment spins will be in a maximally mixed state, that is, $\rho_e=\mathbb{I}_2/2$.

\begin{figure}[t]
\centering
\includegraphics[scale=1.05]{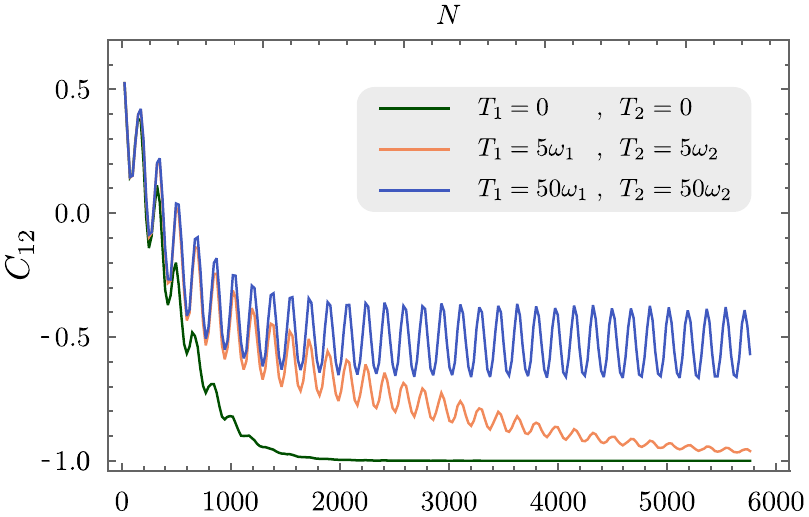}
\caption{Evolution of the Pearson correlation coefficient $C_{12}$ versus the total number of collisions $N$ for three different initial environmental spin pairs $e_1$ and $e_2$ that are thermalized with reservoirs at three different set of equilibrium temperatures, namely, $T_1=0$ and $T_2=0$, $T_1=5 \omega_1$ and $T_2=5 \omega_2$, and $T_1=50 \omega_1$ and $T_2=50 \omega_2$.}
\label{fig6}
\end{figure}

Let us once again take the self-energies of the system spins $s_1$ and $s_2$ to be $\omega_1=1$ and $\omega_2=1.1$ where $\delta t_{s}=0.2$. We also assume that the interaction strength between the system spins are given by $\lambda\delta t_{ss}=0.03$, and both system spins interact with their respective environmental spins $e_1$ and $e_2$ in an identical fashion with the coupling strength $J\delta t_{se}=0.05$. With these considerations in mind, Fig. \ref{fig6} demonstrates the behavior of the Pearson correlation coefficient $C_{12}$ for data windows of 225 collisions with partial overlaps $200$, as a function of the total number of collisions for three different sets of initial environmental spin states corresponding to three different sets of temperatures. Specifically, we show the time evolution of the degree of synchronization for the cases of $T_1=0$ and $T_2=0$, $T_1=5 \omega_1$ and $T_2=5 \omega_2$, and $T_1=50 \omega_1$ and $T_2=50 \omega_2$. As can be easily observed from the plot, increasing equilibrium temperature for the environment spins plays a detrimental role for the emergence of anti-synchronous dynamics between the local observables of the system spins. Or, to state it more plainly, thermal effects on the environmental spins wash away the existence of the phenomenon of spontaneous mutual anti-synchronization.

\section{Conclusion}\label{conclusion}

In summary, we have unveiled the dynamical emergence of spontaneous quantum synchronization between the local observables of a pair of spins in a collision model, where the open system dynamics of the two system spins $s_1$ and $s_2$ are modeled via their brief successive pairwise interactions with a collection of environment spins. Indeed, in our work, what has been observed is the phenomenon of anti-synchronization rather than synchronization. We have performed a comprehensive analysis to fully characterize the conditions under which synchronous dynamics spontaneously occurs between the system spin observables in the absence of an external drive. Our study establishes that the most important ingredient for the manifestation of the environment-induced synchronization is the indirect exchange of information between the system spins $s_1$ and $s_2$, which is mediated by the partial SWAP interaction between the environments spins $e_2$ and $e^{\prime}_1$. 

On the other hand, we have investigated the potential role that is played by the formation of bipartite quantum and total correlations between the system spins $s_1$ and $s_2$, as quantified respectively via concurrence and quantum mutual information, in the dynamical development of synchronization. Making use of the versatile nature of the collision model framework, i.e., having the ability to fully erase the correlations between system spins $s_1$ and $s_2$ after each leg of the collision scheme, we in fact demonstrate that the existence of any sort of correlation between the system spins has no significant connection to the spontaneous manifestation of synchronization provided that the intra-environment couplings are not weak. Furthermore, we have considered an extension of the studied collision model taking into account the consequences of thermal effects on the environment spins. Our findings show that increasing thermal effects on the environmental spins wash away the dynamical emergence of synchronous behavior between the local observables of the system spins.

{Despite the fact that spontaneous mutual synchronization in the quantum domain has been reported in the literature mostly in models in which both the pair of system particles and the environment are described by oscillators, there are a few studies devoted to the  synchronization, where the system particles are considered to be quantum spins such as Ref.~\citep{Giorgi2013}. Our analysis on the other hand demonstrates the manifestation of this phenomenon for the first time in a collision model framework where not only the two system particles are spin objects but also their independent environments are constituted of a large collection of spins.} In closing, we should also mention that we have not considered the possibility of collective interactions of the system spins with more than one environmental spin simultaneously. Thus, it remains as an open question whether the existence of such collective interactions by themselves would be sufficient to induce synchronous dynamics between the system spins even in the absence of intra-environment couplings.

\begin{acknowledgments}

G.K. is supported by the BAGEP Award of the Science Academy and the TUBA-GEBIP Award of the Turkish Academy of Sciences. G.K. and B.\c{C}. are also supported by the Technological Research Council of Turkey (TUBITAK) under Grant No. 117F317. \.{I}.Y.\ is supported by M\v{S}MT under Grant No. RVO 14000. The authors would like to acknowledge the time they spent with Z. Gedik.

\end{acknowledgments}

\bibliography{bibliography}

\onecolumngrid
\appendix
\newpage

\section{Calculation of the Pearson Coefficient}
\label{appendix:A}
\begin{figure}[h]
\centering
\includegraphics[scale=1.7]{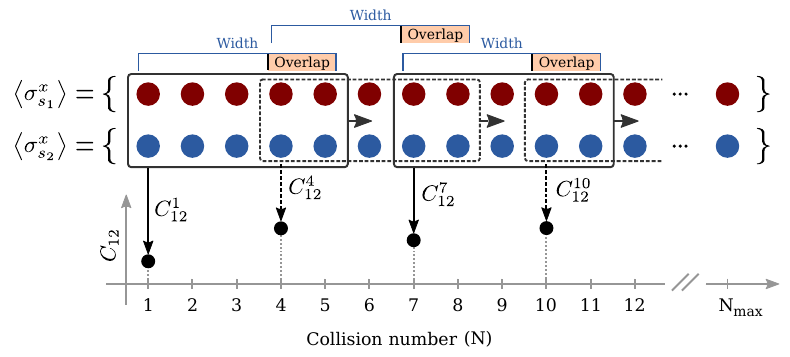}
\caption{{Schematic representation of the evaluation of the Pearson coefficients $C_{12}$ from the numerical data. Red and blue dots represent the data we collect during our numerical simulations for the expectation values of system spin $1$ ($s_1$) and system spin $2$ ($s_2$) along $x$ direction, respectively. Here we choose, in a representative manner, the width of the data window as $5$ and the overlap as $2$. The Pearson coefficient for the first collision, $C_{12}(1)$, is calculated by employing first five of the adjacent data point pairs corresponding to $1$st, $2$nd, $3$rd, $4$th and $5$th collisions, which are covered by the data window. The same calculation is iterated similarly to obtain the other Pearson coefficients, $C_{12}(N)$,  after shifting the data window by multiples of $3$, which is, in this case, $(\text{width})-(\text{overlap})=5-2=3$. Therefore, each $C_{12}(N)$ actually represents the local correlation between $\left<\sigma_{s_1}^x\right>$ and $\left<\sigma_{s_2}^x\right>$ in the right vicinity of $N$th collision where this vicinity is specified by the data window. The overlap specifies the number of data points jointly used while calculating adjacent Pearson coefficients, which is useful to extract more data points for the $C_{12}$ curve, and hence, it smooth outs the resulting curve with an appropriate choice of data window width.}}
\label{fig7}
\end{figure}

{
Our simulations yield two discrete sets of data $\left<\sigma_{s_i}\right>_N$ where $i\in\{1,2\}$ and $N\in\{1,2,\hdots,N_\text{max}\}$ represents number of consecutive collisions. The main target here is to quantify how coherently these two expectation values oscillate with respect to $N$.  The Pearson coefficient $C_{12}$ given in Eq. \eqref{eq:pearsonCoeff} provides a simple tool to serve this purpose. However, one cannot simply calculate it over the whole data set since this would produce a single correlation value, providing an overall correlation between the two data sets. Instead, we calculate many Pearson coefficients along the data set so that each one tracks the local correlation between the expectation values in a restricted data (or equivalently, collision) range specified by a ``data window'' of width $n$ \cite{Galve2017}. In this way, it becomes possible to characterize how a given system switches among synchronous, anti-synchronous and non-synchronous cases throughout its time evolution. This procedure is explained schematically in Fig. \ref{fig7} in detail. At this point, we should mention that there are some important aspects that one needs to pay attention while defining the data window and the overlap. First of all, the data window width has to be chosen to include at least several oscillations since what we seek here is whether these local oscillations are phase-locked or not. If the data width was chosen narrower than a single oscillation, $C_{12}$ would exhibit spikes and thus would not provide any useful information about the synchrony of expectation values. Contrarily, if it was too large, transitions between different behaviors would not be resolved. The overlap, on the other hand, determines the number of data points shared by adjacent data windows and hence it is useful in smoothing out the resulting curve.
}

\section{Effect of the Detuning and Intra-environment Interactions on Synchronization }
\label{appendix:B}

\begin{figure}[h]
\centering
\includegraphics[scale=0.65]{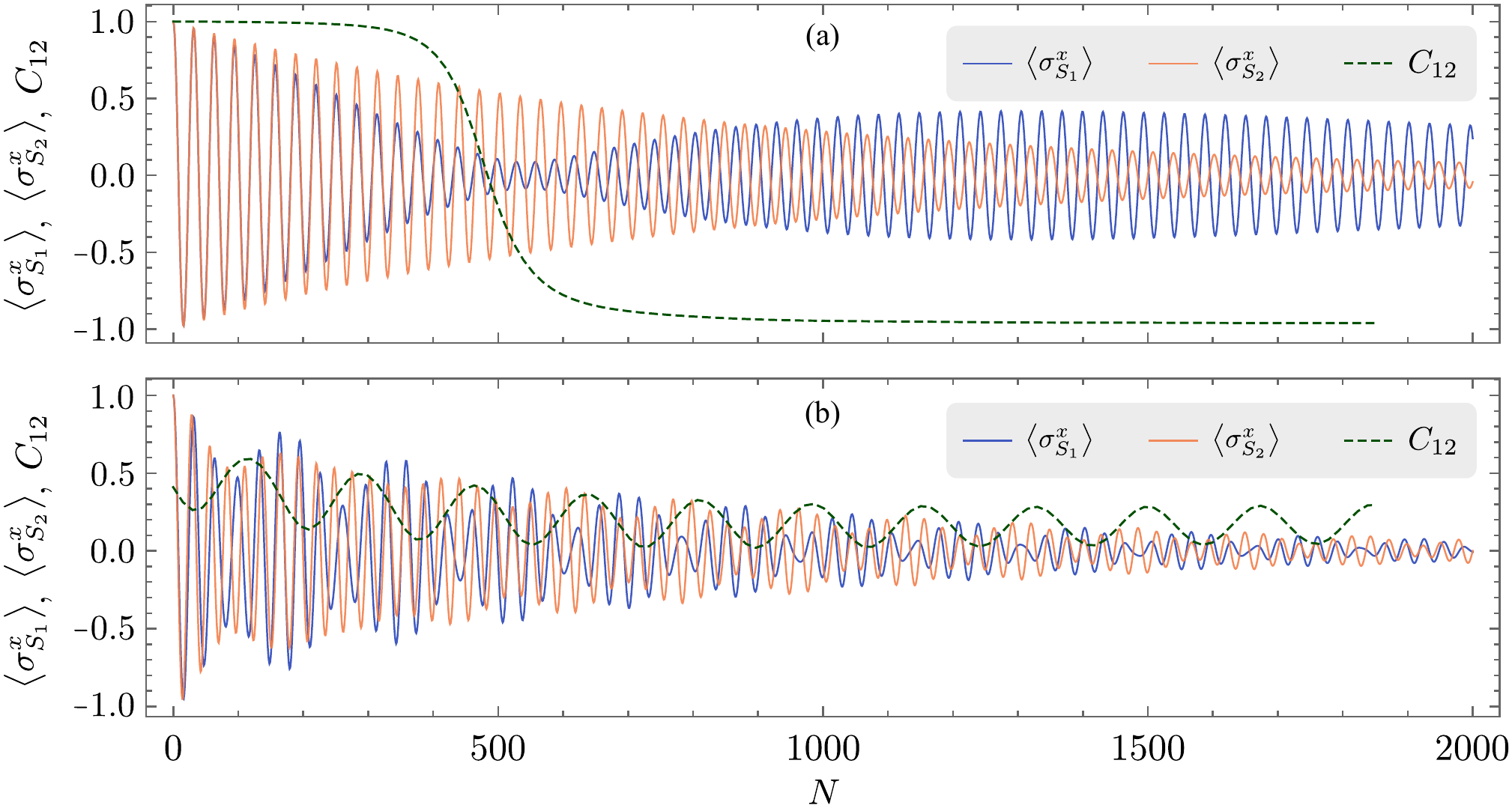}
\caption{{Expectation values for the system spins $s_1$ and $s_2$ along the x-direction and the Pearson  correlation coefficient $C_{12}$, which is plotted for data windows of $140$ collisions with partial overlaps of $125$, versus the total number of collisions $N$. In (a) we suppose that there is no detuning between the self-energies of  $s_1$ and $s_2$, and there exists no direct interaction between them, that is, $\gamma=0.95\pi/2$, $\omega_1=\omega_2$ and $\lambda=0$. In (b) we suppose that $\lambda=0.15$ and $\omega_2=1.1\omega_1$ as in Fig. \ref{fig2} but with $\gamma=0$, i.e., there exists no intra-environment interaction.}}
\label{fig8}
\end{figure}

{Let us first assume that in Fig. \ref{fig8} (a) all the parameters are the same as in the Fig. \ref{fig2} except for the fact that there is no interaction between $s_1$ and $s_2$ and the detuning between them is zero. In this situation, with an identical choice of initial states for the system spins, i.e., $(1/2)(|0\rangle+|1\rangle)\otimes(|0\rangle+|1\rangle)$, we can see from the figure that the spin observables of $s_1$ and $s_2$ evolve trivially in synchrony for a certain number of collisions but then they again dynamically and spontaneously become anti-synchronized after a certain collision interval. On the other hand, in Fig. \ref{fig8} (b) we once again consider the case where all the parameters are the same as in Fig. \ref{fig2} but this time there is no interaction between the environment spins, that is $\gamma=0$. It is easy to see that spontaneous anti-synchronization cannot be established in this case. Hence, the main ingredient for the emergence of anti-synchronization is the indirect transfer of information between $s_1$ and $s_2$, rather than the weak direct interaction between the them, which only accounts for the compensation of the degrading effect of the detuning between their self-energies on anti-synchronization.}

\end{document}